\documentclass[preprint,showpacs,preprintnumbers,amsmath,amssymb]{revtex4}

\usepackage{graphicx}
\usepackage{dcolumn}
\usepackage{bm}

\begin{document}


\title{Topological Discrete Algebra, Ground State Degeneracy, 
and Quark Confinement in QCD}
\author{Masatoshi Sato}
\affiliation{%
The Institute for Solid State Physics, The University of Tokyo,\\
Kashiwanoha 5-1-5,
Kashiwa-shi, Chiba 277-8581, Japan
}%


\date{\today}

\begin{abstract}
Based on the permutation group formalism, we present a discrete symmetry
 algebra in QCD. 
The discrete algebra is hidden symmetry in QCD, which is manifest only
on a space-manifold with non-trivial topology.
Quark confinement in the presence of the dynamical
 quarks is discussed in terms of the discrete symmetry algebra. 
It is shown that the quark deconfinement phase has the ground state
 degeneracy depending on the topology of the space, which gives a
 gauge-invariant distinction between the confinement and deconfinement phases.
We also point out that new quantum numbers relating to the fractional
 quantum Hall effect exist in the deconfinement phase. 
\end{abstract}

\maketitle

The purpose of this paper is to present an argument for classification
of phases in QCD. 
Classification of phases in QCD is an old but unsolved problem in
quantum field theory. (For recent reviews, see
\cite{Greensite03, RW01}.)
As is well-known, behaviors of the Wilson loop \cite{Wilson74} (or
Polyakov line \cite{Polyakov78})
and the 't Hooft loop \cite{tHooft78, tHooft79} are useful to classify
the quark confinement and deconfinement phases in the pure Yang Mills theories,
but once the dynamical quarks are included, they are no longer
sufficient to distinguish them.
Nevertheless, it will be shown below that there exist quantum numbers
which distinguishes these two phases.
In particular, they are useful to study the quark confinement in the
numerical studies.

This work is motivated by recent developments on understanding of 
quantum phases.
Many quantum phases and phase transitions can be described by the
spontaneous symmetry breaking and local order parameters, 
however in recent years it has become increasingly clear that in a wide
class of strongly correlated many-body systems, a phase transition driven by a
nonthermal parameter may occur at zero temperature which can not be
understood by any local order parameter.
The characteristic signature of the novel phase is a finite ground-state
degeneracy depending on the topology of the space,
and the underling order of the novel phase is dubbed as topological order
\cite{Wen90}.
The Laughlin state for the fractional Hall effect is known to have a
topological order \cite{WN90}.  
In the present, many systems including bosonic ones and those at zero
magnetic field are identified as possessing topological orders
\cite{Wen91,WZ91,RS91,SF00,MS01,MSP02,BFG02,MS02,FNSWW04}.

Recently, we have argued that the topological degeneracy in a topological
order is due to the emergence of a discrete symmetry \cite{SKW06}, which
contains three fractional parameters: quasiparticle charge,
anyon statistics, and the fractional quantum Hall conductivity.  
In particular, it is notable that the emergence of collective
excitations having fractional quantum numbers with respect to the
constitute particles in the Hamiltonian is closely related to the
existence of the topological order \cite{SKW06,OS06}. 
Such a charge fractionalization has an interesting analogy of the 
quark deconfinement, which also gives fractional charged excitations. 
In spite of the essential difference
that the quarks are elementary particles, not
collective excitations, 
this motivates us to study the quark confinement in the notion of the
topological order.

In the following, we will show that the quark deconfinement phase in QCD
indeed has a topological order. 
Generalizing the argument in \cite{SKW06},
we will construct a discrete symmetry algebra in QCD, which we dub topological
discrete algebra. 
The existence of the center of the gauge group is crucial for
the construction.
By the use of the topological discrete algebra, it will be shown that the
quark deconfinement phase in QCD has a ground state degeneracy depending on
the topology of the system.
The topological degeneracy in the thermodynamic limit is a
gauge-invariant quantum number that distinguishes the deconfinement phase
from the confinement one. 

For definiteness we will consider the lattice QCD.
The generalization to the continuum QCD is
straightforward. 
The action of the link variable $U_{n,\mu}\in SU(3)$ is given by
\begin{eqnarray}
S_{\rm G}=\sum_{\rm P}\frac{1}{g^2}{\rm tr}(1-U_{\rm P}),
\end{eqnarray}
with the plaquette variable
$
U_{\rm P}=U_{n,\mu}U_{n+\hat{\mu},\nu}
U_{n+\hat{\nu},\mu}^{\dagger}
U_{n,\nu}^{\dagger},
$
and that of the quarks $\psi_{\bm n}^f$ is
\begin{eqnarray}
S_{\rm F}=-\frac{1}{2}\sum_f\sum_{n,\mu}
\left(\bar{\psi}_{n}^f\gamma_{\mu}U_{n,\mu}
\psi_{n+\hat{\mu}}^f-\bar{\psi}_{n+\hat{\mu}}^f
\gamma_{\mu}U_{n,\mu}^{\dagger}\psi_{n}^f\right)
-\sum_{f,{n}}m_f \bar{\psi}_{n}^f\psi_{n}^f,
\end{eqnarray}
where $n=(n_1,n_2,n_3,n_4)$ denotes the site on the lattice,
$\hat{\mu}$ a unit vector in the $n_{\mu}$ direction, and $f$
the indices of the flavors of the quarks.  
To remove the doublers, we also add the Wilson term $S_{\rm W}$.
The partition function ${\cal Z}$ is given by
\begin{eqnarray}
{\cal Z}=\int{\cal D}U{\cal D}\psi{\cal D}\bar{\psi}
e^{-(S_{\rm G}+S_{\rm F}+S_{\rm W})}.
\end{eqnarray}
If necessary, other terms preserving the electric charge (improved
actions for gauge field, the chemical potential terms for quarks, and so
on) can be included, which does not affect the arguments in the following.

\begin{figure}[h]
\begin{center}
\includegraphics[width=3cm]{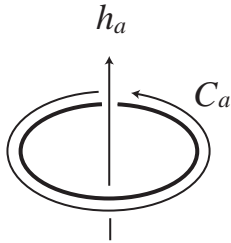}
\caption{Non-contractable loop $C_a$ on $S^1$.}
\label{fig:torus}
\end{center}
\end{figure}

The topological discrete algebra we will consider is defined only
if the topology of the space-manifold is non-trivial,
so let us consider the system on a three dimensional torus $T^3$.
(The space-time is four dimensional.)
On the torus $T^3$, the site at ${\bm n}=(n_1,n_2,n_3)$ is identified
with one at ${\bm n}+\hat{a} N_a$, where $\hat{a}$ is a unit vector in
the $n_a$ direction, and $N_a$ is an integer $(a=1,2,3)$. 
The torus size is $N_1N_2N_3$.
In practice, the torus is realized by imposing the periodic boundary
conditions in all the spatial directions.
The torus admits three independent spatial non-contractable loops $C_a$
$(a=1,2,3)$ along the $n_a$ direction. 
In addition, the torus has three
``holes'' $h_a$ ($a=1,2,3$) outside $T^3$. 
The hole $h_a$ is encircled by the non-contractable loop $C_a$.  
In other words, each $S^1$ in $T^3=S^1\times S^1\times S^1$ has a
noncontractable loop and a hole.
See Fig.\ref{fig:torus}.

To define the topological discrete algebra, we introduce an external $U(1)$
electro-magnetic gauge field $e^{i\theta_{n,\mu}}$.
Then consider an adiabatic insertion of the flux
$\Phi_a$ through  ``hole'' $h_a$ encircled by the
non-contractable loop $C_a$.
This process induces
the external $U(1)$ electro-magnetic gauge field $e^{i\theta_{n,\mu}}$ with
$\theta_{n,\mu}=\delta_{\mu,a}\Phi_a/N_a$.
Note that the $U(1)$ field strength on $T^3$ remains to be zero after the flux
insertion. 

Let us now consider 
the partition function ${\cal Z}(\Phi_a)$ with the inserted flux
$\Phi_a$.
Since the upper and down quarks have $2/3$ and $-1/3$ electric charges,
respectively, then by the unitary transformation, 
\begin{eqnarray}
&&\psi_n\rightarrow e^{-i2n_a\Phi_a/3N_a}\psi_n,
\quad \mbox{for upper quarks},
\nonumber\\
&&\psi_n\rightarrow e^{in_a\Phi_a/3N_a}\psi_n,
\quad \mbox{for down quarks},
\label{eq:quark_trans}
\end{eqnarray}
the induced $U(1)$ electromagnetic gauge field is eliminated
in the action except on the links connecting the sites ${\bm n}$ 
with $n_a=1$ and $n_a=N_a$. 
After the transformation (\ref{eq:quark_trans}), the kinetic terms of
the quarks on these boundaries acquire the $U(1)$ phase as
\begin{eqnarray}
&&e^{i2\Phi_a/3}\bar{\psi}_n \gamma_a U_{n,a}\psi_{n+\hat{a}},
\quad \mbox{for upper quarks},
\nonumber\\
&&e^{-i\Phi_a/3}\bar{\psi}_n \gamma_a U_{n,a}\psi_{n+\hat{a}},
\quad \mbox{for down quarks},
\quad (n_a=N_a).
\end{eqnarray}
The same $U(1)$ phase remains in the kinetic term in the Wilson term.
If $\Phi_a$ is a unit flux $2\pi$, these $U(1)$ phases coincide with an
element of the
center of $SU(3)$, $e^{-2\pi i/3}$. 
So performing the unitary transformation on link variables $U_{n,a}$ on
the boundaries,
$
U_{n,a}\rightarrow e^{2\pi i /3}U_{n,a},
(n_a=N_a)
$,
one can show  
$
{\cal Z}(2\pi)={\cal Z}(0)
$.
Therefore, the spectrum of the system is invariant under the adiabatic
flux insertion by $2\pi$.
The unit flux $2\pi$ is smaller than that in \cite{SKW06} where the unit
flux is $2\pi/e$ with $e$ the minimal charge of the constituent
particles. 
In QCD, $e=1/3$. 
The adiabatic insertion of the unit flux $2\pi$ is represented by a unitary
operator $U_a$.

If quarks are deconfined, the physical states
are classified by the representation of the permutation group
for quarks. 
For $N$ quarks on $T^3$, 
the permutation group consists of $\sigma_i$ $(i=1,\cdots, N-1)$, which
exchanges the $i$th and $(i+1)$th quarks clockwise without
enclosing any other quark, and $\tau_i^{a}$
$(a=1,2,3, \, i=1,\cdots,N)$, which represents moving the $i$th
quark along the non-contractable loop $C_a$ in
the $n_a$ direction. 
The permutation group is given by 
\begin{eqnarray}
\sigma_k^2=1,
\quad 1\le k \le N-1,
\nonumber\\
(\sigma_k\sigma_{k+1})^3=1,
\quad
1\le k \le N-2,
\nonumber\\ 
\sigma_k\sigma_l=\sigma_l\sigma_k, 
\quad
1\le k \le N-3, \quad |l-k|\ge 2,
\nonumber\\
\tau_{i+1}^a=\sigma_i \tau_i^{a}\sigma_i,
\quad
1\le i \le N-1,
\quad a=1,2,3,
\nonumber\\
\tau_1^{a}\sigma_j =\sigma_j \tau_1^{a},
\quad
2\le j \le N, 
\quad
a=1,2, 3,
\nonumber\\
\tau_i^{a}\tau_j^{b}=\tau_j^{b}\tau_i^a,
\quad
i,j=1,\cdots, N,
\quad
a,b=1,2, 3.
\label{eq:permutation}
\end{eqnarray}
The generators of the permutation group
have non-trivial commutation relations with $U_a$ $(a=1,2,3)$. 
If we apply $\tau_i^a$ after the flux insertion $U_a$, the gauge field
$e^{i\theta_{n,\mu}}$ will give rise to an 
Aharanov-Bohm phase $e^{-i2\pi/3}$. (Both the upper and down quarks
acquire the same $U(1)$ phase.)
Therefore, we obtain
\begin{eqnarray}
\tau_i^{a}U_a=e^{-2\pi i/3}U_a\tau_i^{a},
\label{eq:topological_algebra1}
\end{eqnarray}
where $a=1,2,3$ and $i=1,\cdots,N$.
On the other hand, because $\tau_i^{b}$ $(b\neq a)$ and $\sigma_i$ do not
encircle the inserted flux by $U_a$, they commute with $U_a$, 
\begin{eqnarray}
\tau_i^{b}U_a=U_a\tau_i^{b},
\quad
\sigma_i U_a =U_a\sigma_i.
\label{eq:topological_algebra2}
\end{eqnarray}
with $a\neq b$.

Using the commutation relations (\ref{eq:topological_algebra1}) and
(\ref{eq:topological_algebra2}), one can verify that $U_aU_b U_a^{-1}
U_b^{-1}$ $(a\neq b)$ commutes with all the permutation group generators. 
So by Schur's lemma, for any irreducible representation of the
permutation group,
$U_aU_bU_a^{-1}U_b^{-1}$ is a (unimodular) $c$-number.
Namely,
\begin{eqnarray}
U_a U_b=e^{2\pi i \lambda_{a,b}}U_b U_a, 
\quad
\lambda_{a,b}=-\lambda_{b,a}.
\label{eq:UU}
\end{eqnarray}
In addition, $U_a^3$ is also a unimodular integer because it also
commutes with all the permutation group generators. 
Therefore $U_a^3U_b=U_bU_a^3$. 
Comparing this with (\ref{eq:UU}), we find that $\lambda_{a,b}$
should be a rational number 
\begin{eqnarray}
\lambda_{a,b}=\frac{k_{a,b}}{3}, 
\end{eqnarray}
where $k_{a,b}$ is an integer and co-prime to 3.
These constants $\lambda_{a,b}$ $(a,b=1,2,3)$ are new quantum
numbers in the deconfinement phase.

If the system is time-reversal invariant, these new quantum numbers are
shown to be zero:
For the time-reversal invariant system, the time-reversal
transformation of $U_a$ is given by
$
T U_a T^{-1}=c_a U_a^{\dagger}
$
with a unimodular constant $c_a$.
Applying the time-reversal transformation $T$ to (\ref{eq:UU}) and using
the anti-Hermiticity of $T$, we have 
\begin{eqnarray}
U_a^{\dagger}U_b^{\dagger}=e^{-2\pi i \lambda_{a,b}}U_b^{\dagger} 
U_a^{\dagger}. 
\label{eq:UdUd}
\end{eqnarray}
The compatibility between (\ref{eq:UU}) and (\ref{eq:UdUd}) leads to 
$\lambda_{a,b}=0$.

In 3+1 dimensions, the only allowed statistics for particle excitations
is boson or fermion, 
$\sigma_i=\pm {\bm 1}$. 
Then the permutation group representation is uniquely determined as
$
\tau_i^a=T_a
$,
with matrices $T_a$ satisfying 
\begin{eqnarray}
T_a T_b=T_b T_a. 
\label{eq:topological_algebra_deconfinement1}
\end{eqnarray}
The commutation relations (\ref{eq:topological_algebra1}),
(\ref{eq:topological_algebra2}) and (\ref{eq:UU}) now reduce to
\begin{eqnarray}
T_a U_b = e^{-(2\pi i/3)\delta_{a,b}} U_b T_a,
\quad
U_a U_b=e^{2\pi i \lambda_{a,b}}U_b U_a, 
\label{eq:topological_algebra_deconfinement2}
\end{eqnarray}
with $a,b=1,2,3$.
The topological discrete algebra in the quark deconfinement phase consists of
the flux insertion operators $U_a$ and the quark winding operators $T_a$
with the commutation relations
(\ref{eq:topological_algebra_deconfinement1}) and
(\ref{eq:topological_algebra_deconfinement2}).

Now we show our main claim in this paper: {\it If the ground state of
QCD has a mass gap, the quark deconfinement phase has at least $3^3$-fold 
ground state degeneracy on $T^3$.} 
To show this, consider the following process.   
First, create a pair of quark and anti-quark out of a vacuum, then
move the quark by $T_a$.
After the quark returns to the original position, we pair annihilate the quark
and the anti-quark.
Suppose that there is a mass gap to excitations above the vacuum space and
these operations do not close the mass gap, 
then these processes define the operation of $T_a$ from a vacuum to
a vacuum.
Since $T_a$ $(a=1,2,3)$ commutes with each other, we can take the
basis of the vacuum space which diagonalizes $T_1$, $T_2$ and $T_3$
simultaneously,
$
T_a|{\bm \eta}\rangle = e^{i\eta_a}|{\bm \eta}\rangle, 
{\bm \eta}=(\eta_1,\eta_2,\eta_3)
$.
By applying $U_a$ $(a=1,2,3)$ to this and using
(\ref{eq:topological_algebra_deconfinement2}), we have
\begin{eqnarray}
T_1\left(U_1^r U_2^s U_3^t|{\bm \eta}\rangle\right)
=e^{i(\eta_1-2\pi r/3)}
U_1^r U_2^s U_3^t|{\bm \eta}\rangle,
\nonumber\\
T_2\left(U_1^r U_2^s U_3^t|{\bm \eta}\rangle\right)
=e^{i(\eta_2-2\pi s/3)}
U_1^r U_2^s U_3^t|{\bm \eta}\rangle,
\nonumber\\
T_3\left(U_1^r U_2^s U_3^t|{\bm \eta}\rangle\right)
=e^{i(\eta_3-2\pi t/3)}
U_1^r U_2^s U_3^t|{\bm \eta}\rangle,
\end{eqnarray}
where $r$, $s$ and $t$ are integers. 
Therefore, it is found that there are $3^3$ distinct sets of eigenvalues
of $T_a$'s.
This implies that the ground state (vacuum) in the quark deconfinement
phase has at least $3^3$-fold degeneracy.

On the other hand, if quarks are confined, the topological discrete algebra
becomes trivial and the ground state degeneracy
obtained above disappears:
In the confinement phase, 
the permutation group for hadrons, not for quarks, 
classifies the physical states.
The permutation group for hadrons is also defined by
(\ref{eq:permutation}) if $\sigma_i$ and $\tau_i^a$ are
interpreted as those for hadrons.
On the contrary to the quark deconfinement phase, however, all the
generators of the permutation group for hadrons commute with the flux insertion
operators $U_a$ $(a=1,2.3)$. 
This is because hadrons have integer electric charges. 
From this property, the movement $\tau_i^a$ of a hadron around the
inserted flux $\Phi_a=2\pi$ gives only the trivial Aharanov-Bohm phase, which
leads to 
$
\tau_i^a U_b=U_b \tau_i^a
$,
$
\sigma_iU_a=U_a\sigma_i,
(a,b=1,2,3)
$.
Then by the Schur's lemma, the flux insertion operator $U_a$
reduces to a unimodular constant for any irreducible representation of
the permutation group for hadrons.
In addition, since the allowed representations of the permutation group
for hadrons are fermion and boson, 
$\tau_i^a$ for a hadron is again uniquely determined as
$\tau_i^a=\tilde{T}_a$ with mutually commuting matrices  $\tilde{T}_a$
$(a=1,2,3)$. 
Because all the elements of the topological discrete algebra commute with each
other, no ground state degeneracy is obtained in the confinement phase. 

Note that the ground state degeneracy obtained in the quark deconfinement phase
depends on the topology of the space-manifold.
This is easily seen by considering the system on a 3-dimensional
box with the free boundary conditions, which is homotopic to the
3-dimensional ball $B^3$.
Because no non-contractable loop exists on this space-manifold, the
permutation group consists of only the exchange operators
$\sigma_{i}$, and does not include $\tau_i^a$. 
Moreover, the operators $U_a$ $(a=1,2,3)$ can not be defined.
So no topological discrete algebra is defined and no ground state degeneracy is
obtained in this space-manifold. 
In general, if the space-manifold on which the system is defined has $l$
independent spatial non-contractable loops, then the minimal ground state
degeneracy in the deconfinement phase becomes $3^l$.
Our results here indicate that the deconfinement
phase in QCD is topologically ordered, and the quark confinement and
deconfinement transition is described properly by the concept of topological
order.

In the static limit of QCD, the minimal topological degeneracy obtained
above is reproduced 
by a conventional argument using the Wilson loop along the
non-contractable loop $C_a$, 
$
W(C_a)={\rm tr} \prod_{n \in C_a} U_{n,a}
$.
In this limit, all the quarks are infinitely heavy and
decoupled from the dynamics.
So the system is effectively described by the
pure $SU(3)$ gauge theory.
The pure $SU(3)$ gauge theory is invariant under
the transformations,
\begin{eqnarray}
U_{n,a}\rightarrow e^{2 i \pi m /3}U_{n,a},
\quad (m=1,2,3),
\quad \mbox{$n_a$ fixed}, 
\end{eqnarray}
which rotate all spacelike links in the $n_a$ direction at a fixed $n_a$
by an element of the center of $SU(3)$.
The Wilson loop $W(C_a)$ is transformed by this center symmetry as
$
W(C_a)\rightarrow e^{i2m\pi/3} W(C_a)
$,
so the expectation value $\langle W(C_a)\rangle$ is an order parameter
for the center symmetry.
In the quark confinement phase, from the area law, it follows that in the
temporal gauge
$
\langle W(C_a, \tau) W^{\dagger}(C_a, \tau')\rangle \sim 
e^{-\sigma N_a|\tau-\tau'|}, 
$
with the imaginary times $\tau=it$ and $\tau'=it'$.
Here $\sigma$ is a positive constant.
Thus using the cluster property
\begin{eqnarray}
\langle W(C_a, \tau) W^{\dagger}(C_a, \tau')\rangle 
\stackrel{|\tau-\tau'|\rightarrow \infty}{\longrightarrow}   
|\langle W(C_a)\rangle|^2,
\end{eqnarray}
we have 
$
\langle W(C_a) \rangle=0
$,
so the center symmetry is not broken.
On the other hand, in the quark deconfinement phase, it is possible that
$
\langle W(C_a)\rangle \neq 0
$,
and the center symmetry can be spontaneously broken.
If $\langle W(C_a)\rangle\neq 0$ for all $C_a$'s $(a=1,2,3)$, 
we have $3^3$ different set of $\langle W(C_a)\rangle$'s, which are
related to each other by the center symmetry.
Thus the ground state degeneracy is $3^3$-fold, and 
it coincides with the minimal ground state degeneracy obtained from the
topological discrete algebra.

The topological discrete algebra constructed above has a similarity to the 't
Hooft algebra \cite{tHooft78, tHooft79}.
For example our relation
\begin{eqnarray}
T_a U_b =e^{-(2\pi i/3)\delta_{a,b}}U_b T_a 
\end{eqnarray}
corresponds to the following relation given by the 't Hooft, 
\begin{eqnarray}
W(C)B(C')=B(C')W(C)e^ {2\pi i n/{\cal N}_{\rm c}},
\quad
{\cal N}_{\rm c}=3
\label{eq:thooft}
\end{eqnarray}
where $C$ and $C'$ denote closed curves in 3-dimensional space,
$n$ the number of times the curve $C'$ winds around $C$ in a
certain direction, and $B(C')$ the 't Hooft loop along $C'$.
However, there exist essential distinctions between them.
First of all, the 't Hooft algebra can be defined only when the dynamical
quarks are absent, but our topological discrete algebra is defined in the
presence of the dynamical quarks.
Second, the topological discrete algebra in the quark deconfinement phase is
different from that in the quark confinement phase, but
the 't Hooft algebra  is the same in any phases. 
Third, new quantum numbers $\lambda_{a,b}$, which are missing in the 't
Hooft algebra, exist in the topological discrete algebra.

Now we would like to address the physical meaning of
 $\lambda_{a,b}$.
Consider the degenerate ground states $\phi_K=|{\bm \Phi}\rangle_K$
 $(K=1,\cdots,d)$
in the presence of inserted fluxes ${\bm \Phi}=(\Phi_1,\Phi_2,\Phi_3)$. 
They satisfy
$
U_a|{\bm \Phi}\rangle_K=e^{i\gamma_a({\bm \Phi})}
|{\bm \Phi}+\hat{a} 2\pi \rangle_K
$,
where $\gamma_a(\Phi_a)$ is the quantum holonomy given by
\begin{eqnarray}
\gamma_a({\bm \Phi})=i\int_{\Phi_a}^{\Phi_a+2\pi}
d\Phi_a \langle \phi_K|\frac{\partial}{\partial \Phi_a}
|\phi_K\rangle. 
\end{eqnarray}
From $U_aU_b=e^{2\pi i\lambda_{a,b}}U_bU_a$, it follows
\begin{eqnarray}
\gamma_a({\bm \Phi}+\hat{b}2\pi)+\gamma_b({\bm \Phi})-\gamma_b({\bm \Phi}
+\hat{a}2\pi)-\gamma_a({\bm \Phi})=2\pi \lambda_{a,b}+2\pi M,
\end{eqnarray}
where $M$ is an integer,
which is independent of $K$ because in general the degenerate ground
states are related to each other by some symmetry.
Then by the Stokes's theorem, the Hall conductance $\sigma_{ab}$ $(a\neq
b)$ \cite{NTW85}
\begin{eqnarray}
\frac{e^2}{hd}\sum_{K=1}^{d}\int_0^{2\pi}\int_{0}^{2\pi}
\frac{d\Phi_a d\Phi_b}{2\pi i}
\left[\left\langle\frac{\partial \phi_K}{\partial\Phi_a}| 
\frac{\partial \phi_K}{\partial \Phi_b}\right\rangle
-(\Phi_a\leftrightarrow\Phi_b) \right]
\end{eqnarray}
reduces to 
\begin{eqnarray}
\sigma_{ab}=\frac{e^2}{h}(\lambda_{a,b}+M). 
\end{eqnarray}
This indicates clearly that a fractional $\lambda_{a,b}$ implies 
the fractional quantum Hall effect.

The above discussions can be generalized to the $SU({\cal N}_{\rm c})$ QCD.
For this purpose, it is convenient to introduce a
fictitious $U(1)$ gauge field as an external field, instead of the
electromagnetic gauge field, and assign the fictitious $U(1)$ charge
${\cal Q}=1/{\cal N}_{\rm c}$ to all quarks. 
Then, consider an adiabatic insertion of the fictitious $U(1)$ flux by
$2\pi$ through the hole $h_a$ of $T^3$,
which is represented by a unitary operator ${\cal U}_a$ $(a=1,2,3)$.  
The quark winding operator $T_a$ is defined in the same way as above.
The topological discrete algebra in the quark deconfinement phase is given by
\begin{eqnarray}
T_a T_b=T_b T_a,
\quad
T_a {\cal U}_b=e^{(2\pi i/{\cal N}_{\rm c})\delta_{a,b}}{\cal U}_bT_a,
\quad
{\cal U}_a {\cal U}_b=e^{2\pi i k_{a,b}/l_{a,b}}{\cal U}_b {\cal U}_a,
\end{eqnarray}
where $k_{a,b}$ and $l_{a,b}$ are co-prime integers,
and $l_{a,b}$ is a divisor of ${\cal N}_{\rm c}$.
Using these relations, the minimal ground state degeneracy on
$T^3$ in the quark deconfinement phase
is found to be ${\cal N}_{\rm c}^3$.
On the other hand, in the quark confinement phase, the topological discrete
algebra is trivial and no topological degeneracy arises because the
charge ${\cal Q}$ of any hadron is an integer.

To count the ground state degeneracy, we have assumed that the system
has a finite gap. 
While color charges are screened in the presence of the gluon mass gap, our
result indicates that the quark confinement is not synonymous with the
color screening.   
In addition, the topological discrete algebra itself is valid even in the
gapless system. 
The concept of ground state degeneracy becomes subtle in the gapless
system, but the degeneracy could be identified by examining the finite
scaling carefully.

To conclude, we have argued phases in QCD by using a discrete
symmetry algebra which is manifest only on a space with non-trivial
topology. 
The topological degeneracy of the ground state, which indicate the
presence of a topological order, is derived and it is found that even in
the presence of dynamical quarks it is a good quantum number
distinguishing the quark confinement phase from the deconfinement one.

The author would like to thank H. Kawamura for helpful discussions.

\bibliography{topological_order}

\end{document}